\documentclass[conference]{IEEEtran}
\usepackage{ifpdf}

\usepackage{cite}

\usepackage{graphicx}
\graphicspath{{fig}{images}}
\DeclareGraphicsExtensions{.ps,.eps,.pdf,.jpeg,.png}

\usepackage[lineno5,leftno]{lgrind}
\usepackage{algorithmic}
\begin{document}
%
\title{Parallel Sparse Matrix Solver on the GPU Applied to Simulation of Electrical Machines}

\author{\IEEEauthorblockN{Wendell O. Rodrigues}
\IEEEauthorblockA{LIFL - USTL\\\small{INRIA Lille Nord Europe} - 59650\\Villeneuve d'Ascq - France\\wendell.rodrigues@inria.fr}
\and
\IEEEauthorblockN{Fr\'{e}d\'{e}ric Guyomarc'h}
\IEEEauthorblockA{LIFL - USTL\\\small{INRIA Lille Nord Europe} - 59650\\Villeneuve d'Ascq - France\\frederic.guyomarch@inria.fr}
\and
\IEEEauthorblockN{Yvonnick Le Menach}
\IEEEauthorblockA{L2EP USTL\\\small{Cit\'{e} Scientifique Bat.P2 - 59655}\\Villeneuve d'Ascq - France\\yvonnick.le-menach@univ-lille1.fr}
\and
\IEEEauthorblockN{Jean-Luc Dekeyser}
\IEEEauthorblockA{LIFL - USTL\\\small{INRIA Lille Nord Europe} - 59650\\Villeneuve d'Ascq - France\\jean-luc.dekeyser@lifl.fr}
}


%


\maketitle

\begin{abstract}
Nowadays, several industrial applications are being ported to parallel architectures. In fact, these platforms allow acquire more performance for system modelling and simulation. In the electric machines area, there are many problems which need speed-up on their solution. This paper examines the parallelism of sparse matrix solver on the graphics processors. More specifically, we implement the conjugate gradient technique with input matrix stored in CSR, and Symmetric CSR and CSC formats. This method is one of the most efficient iterative methods available for solving the finite-element basis functions of Maxwell's equations. The GPU (Graphics Processing Unit), which is used for its implementation, provides mechanisms to parallel the algorithm. Thus, it increases significantly the computation speed in relation to serial code on CPU based systems. 
\end{abstract}


%
\IEEEpeerreviewmaketitle

\section{Introduction}
An electromagnetic field analysis is one of the most complex problems of physics. And, the modeling and simulation of electrical systems use a very large computational algorithms to solve their problems. One of this algorithms is based on the conjugate gradient (CG) method \cite{cg}. The CG method is an effective technique for symmetric positive definite systems. It is suitable for systems of the form \emph{Ax=b}, where \emph{A} is a known, square, symmetric, positive-definite matrix, \emph{x} is a unknown solution vector and \emph{b} is a known vector. Iterative methods like CG are suited for use with sparse matrices.

The solution of large, sparse linear systems of equations is the single most computationally expensive step. Thus, any reduction in the linear system solution time will result in a significant saving in the total process time. This need demands for algorithms and software that can be used on parallel processors.

This paper describes the use of the GPU as platform to implement a CG method to solve a linear system. The sample inputs are matrix stored in Symmetric CSR (compressed sparse row)\cite{cg} format obtained from a simulation of electric machines. The algorithms use standard BLAS library and two implemented kernels for matrix-vector product (SpMV). The next section describes the GPU which is the platform used and how to program it with CUDA. Section 3 depicts the matrix used and how to implement the solver. A execution performance is discussed in following section. Finally, some concluding remarks are made about obtained results.

\section{GPU}
GPU is a manycore processor attached to a graphics card dedicated to calculating floating point operations. Even if GPUs are a manycore processors, their parallelism continues to scale with Moore's law. It is necessary to develop application software that transparently scales its parallelism. CUDA (Compute Unified Device Architecture)\cite{cuda} is a parallel programming model and software environment designed to overcome this challenge while maintaining a low learning curve for programmers familiar with standard programming languages such as C.

 

\subsection{Processor Architecture}
The GPU devotes more transistors to data processing rather data caching and flow control. This is the reason why the GPU is specialized for compute intensive. NVIDIA GPU, more precisely,
 is composed of array of SM(Streaming Multiprocessors), each one is equipped with 8 scalar cores (the SP or Streaming Processors), 16834 32-bit registers, and 16KB of high-bandwidth low-latency memory shared for up to 1024 co-resident threads. GPUs such as the NVIDIA GeForce GTX 280 contain 30 multiprocessors, so 30K threads can be created for a certain task. Further, each multiprocessor executes groups, called \emph{warps}, of 32 threads simultaneously.


\subsection{Memory Architecture}
In the NVIDIA GPU memory model, there are per-thread local, per-block shared, and device memory which comprehends global, constant, and texture memories. Shared Memory can be only accessed by threads in the same block. Because it is on chip, the Shared Memory space is much faster than the local and Global Memory spaces. But only 16KB of shared memory are available on each SM.

\subsection{CUDA Programming Model}
CUDA is a C language extension developed by NVIDIA to facilitate writing programs on GPUs. It allows the programmer to define C functions, called \emph{kernels}, that, when called, are executed N times in parallel by N different \emph{CUDA threads}, as opposed to only once like regular C functions. One of the main features of CUDA is the provision of a Linear Algebra library(CuBLAS) and an Fast Fourier Transform library (CuFFT) \cite{cuda}. The next section describes the implementation of CG on the GPU and the use of CuBLAS library.

\section{Implementation}
The Conjugate Gradient was applied to 30880x30880 symmetric sparse matrix $A$ with 449798 nonzero double precision elements. It is stored in full or symmetric CSR format files which include the unknown elements vector \emph{x} and the result vector \emph{b}. The CG algorithm used to implement the program is below.
\hfill
\begin{algorithmic}[1]
\REQUIRE init variables
\FORALL{$k$ such that $1\leq k\leq N$}
\STATE $\alpha \Leftarrow \frac{(r^\mathrm{T} r)}{(p^\mathrm{T} A  p)}$
\STATE $x \Leftarrow x + \alpha p$
\STATE \emph{break} \textbf{if} \emph{convergence}
\STATE $r \Leftarrow r - \alpha A p$
\STATE $\beta \Leftarrow \frac{(r_{k}^\mathrm{T} r_{k})}{(r_{k-1}^\mathrm{T} r_{k-1})}$
\STATE $p \Leftarrow r + \beta p$
\ENDFOR
\end{algorithmic}

All the internal loop operations are executed on GPU. The scalar product and \emph{axpy} functions on lines 3,5,7 use  CuBLAS functions. Since BLAS is only for dense matrix, naturally it is necessary to create a \emph{CUDA kernel} for sparse matrix. For that reason, a SpMV algorithm for matrix-vector multiplication should be implemented to execute the $A p$ product on line 2. Let $A=(L+D+U)$, where $D$ is the main diagonal part of $A$, $L$ is its strictly lower triangular part and $U$ is its strictly upper triangular part. Since $A$ is symmetric, $U^\mathrm{T}=L$. Thus, $A$ can be stored in CSR format of $(L+D)$. 

\begin{figure}[!t]
\centering
\includegraphics[width=3in]{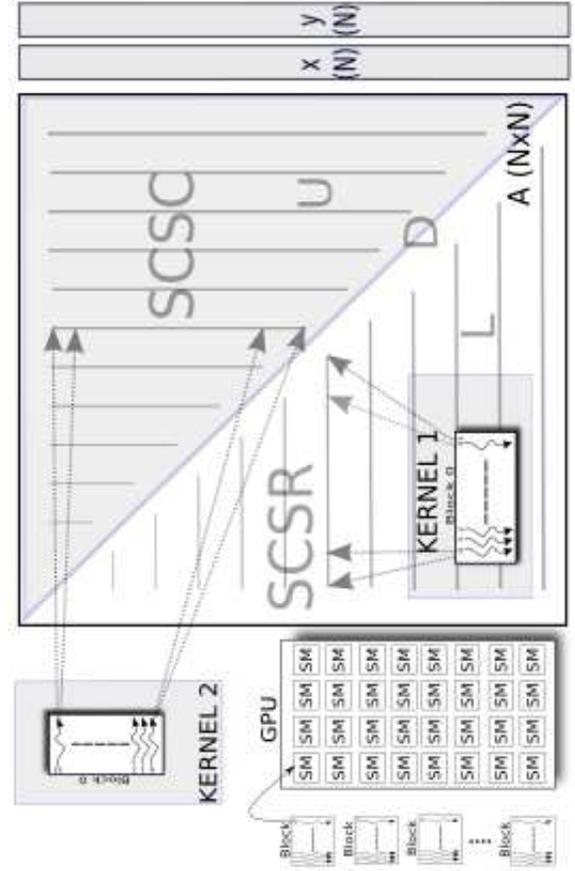}
\caption{SpMV on Symmetric CSR and CSC storage format}
\label{fig_algo}
\end{figure}

Three algorithms were implemented in this work. The first one is a trivial solution of $y=Ax$ in which each thread executes $y[tid]+= A[i-1]*x[jA[i-1]-1]$, where \emph{tid} is the thread identify, and  [\emph{i},\emph{jA}] are obtained from vectors of the stored matrix in compressed form. The two other solutions explore the symmetry characteristic. The aim is to cut down the time and memory allocation cost. The figure \ref{fig_algo} shows the procedure of algorithm conception. The \emph{kernels} are composed of blocks and each block has certain threads. Only one kernel can execute in same time on same device and there are a few blocks in execution simultaneously in each SM. Sequentially the \emph{kernel} 1(SCSR) and \emph{kernel} 2(SCSC) are executed and they take care of $L+D$ and $U$ respectively. To avoid writing conflicts on $y$ vector  in global memory, the \emph{kernel} 2 calls \emph{atomic functions} and that allows just one thread write in a memory address at given moment. As the number of elements of a column or a row can be greater than the number of threads in a block, it is necessary that the algorithm calculates the quantity of elements for each thread. 

\section{Results}
A double precision version of CG was created on a contemporary conventional CPU and this one was used as a reference to calculate the speed-up. Two other versions were developed on the GPU. The first one is a single precision version which was tested on a NVIDIA Tesla C870 card. The second one is a double precision version which was executed on a NVIDIA GeForce GTX 280.

The matrix reading from files takes time to be done properly. Approximately 6ms in our tests. Because this operation is basic for all experiments, it is not necessary to evaluate it. It is important to examine the \emph{Scalar Product, AXPY, and SpMV} execution time because these functions are called repetitively in each loop interaction. Yet, all the time that the interactive CG loop takes. The table \ref{table1} shows the operations and their respective spent time values. The application not use the \emph{sym SpMV}, its times are there just for benchmarking.
\begin{table}[h!b!p!]
\centering
\caption{Speed-Up Results}
\begin{tabular}{r|rrr}
\hline\\[-1em]
Operation & CPU Time & Single Prec. C870 & Double Prec. GTX280 \\
\hline\\[-1em]
dotProd & 0.290ms & 0.529ms & 0.450ms \\
AXPY & 0.109ms & 0.014ms & 0.008ms \\
SpMV & 5.966ms &  0.013ms & 0.005ms \\
SpMV(sym) & - &  0.040ms & 0.035ms \\
\hline\\[-1em]
CG/ \# int & 2392ms/328 & 400ms(5.8x)/324 & 370ms(6.4x)/323 \\
\hline
\end{tabular}
\label{table1}
\end{table}

\section{Conclusion}
This approach issues the viability of using GPU on sparse matrix solvers. The obtained results allow us to evaluate two important impacts in finite-element method of Maxwell's equations using Conjugate Gradient algorithm. The first one is concerned to speed gain. On the GPU, the CG arrives faster than on the CPU due its parallel architecture. The second one is the impact of storing symmetric matrix in memory. Although it can store in memory almost half of the matrix, the performance of algorithm decreases as shown in table of speed-up results. This is due the repetitive way that the \emph{kernel} computes and accumulates the $y=Ax$ result. However, both cases contribute significantly to accelerate the computation of basis functions.






%

\end{document}